\documentclass[aps, longbibliography, amsmath,amssymb, superscriptaddress, 12pt]{revtex4-1}

\usepackage{graphicx}
\usepackage{dcolumn}
\usepackage{bm}
\usepackage{lineno}

\usepackage{color,graphicx,subfigure}
\usepackage{enumitem}
\usepackage{wrapfig}
\usepackage{soul}
\usepackage{amsmath}
\usepackage{titlesec}
\usepackage{multirow}
\usepackage{textcomp}
\usepackage{gensymb}
\usepackage[]{hyperref}
\hypersetup{colorlinks=true,linkcolor=blue,citecolor=blue,urlcolor=blue,pdfpagemode=UseNone}
\usepackage{xcolor}
\usepackage{etoolbox}
\usepackage{floatrow}
\usepackage{floatflt}

\newcommand{\bscco}{$\textrm{Bi}_2\textrm{Sr}_2\textrm{Ca}\textrm{Cu}_2\textrm{O}_{8+\delta}$}

\newcommand{\ybco}{$\textrm{Y}\textrm{Ba}_2\textrm{Cu}_3\textrm{O}_{6+\delta}$}

\newcommand{\qco}{${q}_{CO}$}
\newcommand{\lnsco}{$\textrm{La}_{1.6-x}\textrm{Nd}_{0.4}\textrm{Sr}_x\textrm{Cu}\textrm{O}_{4}$}

\definecolor{eh}{rgb}{1, 0, 0}

\begin{document}

\title{Low-energy quasi-circular electron correlations with charge order wavelength in \bscco}


\author{K.\,Scott}
\affiliation{\footnotesize \mbox{Department of Physics, Yale University, New Haven, Connecticut 06520, USA}}
\affiliation{\footnotesize \mbox{Energy Sciences Institute, Yale University, West Haven, Connecticut 06516, USA}}

\author{E.\,Kisiel}
\affiliation{\footnotesize Department of Physics, University of California San Diego, La Jolla, California 92093, USA}

\author{T.\,J.\,Boyle}
\affiliation{\footnotesize \mbox{Department of Physics, Yale University, New Haven, Connecticut 06520, USA}}
\affiliation{\footnotesize \mbox{Energy Sciences Institute, Yale University, West Haven, Connecticut 06516, USA}}
\affiliation{\footnotesize \mbox{Department of Physics and Astronomy, University of California, Davis, California 95616, USA}}

\author{R.\,Basak}
\affiliation{\footnotesize Department of Physics, University of California San Diego, La Jolla, California 92093, USA}

\author{G.\,Jargot}
\affiliation{\footnotesize Centre \'{E}nergie Mat\'{e}riaux T\'{e}l\'{e}communications, Institut National de la Recherche Scientifique, Varennes, Qu\'{e}bec J3X 1S2, Canada}


\author{S.\,Das}
\affiliation{\footnotesize Department of Physics, University of California San Diego, La Jolla, California 92093, USA}

\author{S.\,Agrestini}
\affiliation{\footnotesize Diamond Light Source, Harwell Campus, Didcot OX11 0DE, United Kingdom}

\author{M.\,Garcia-Fernandez}
\affiliation{\footnotesize Diamond Light Source, Harwell Campus, Didcot OX11 0DE, United Kingdom}

\author{J.\,Choi}
\affiliation{\footnotesize Diamond Light Source, Harwell Campus, Didcot OX11 0DE, United Kingdom}

\author{J.\,Pelliciari}
\affiliation{\footnotesize \mbox{National Synchrotron Light Source II,
Brookhaven National Laboratory, Upton, NY 11973, USA}}

\author{J.\,Li}
\affiliation{\footnotesize \mbox{National Synchrotron Light Source II,
Brookhaven National Laboratory, Upton, NY 11973, USA}}

\author{Y.\,D.\,Chuang}
\affiliation{\footnotesize \mbox{Advanced Light Source, Lawrence Berkeley National Laboratory, Berkeley, CA 94720, USA}}

\author{R.\,D.\,Zhong}
\affiliation{\footnotesize \mbox{Condensed Matter Physics and Materials Science, Brookhaven National Laboratory, Upton, NY, USA}}

\author{J.\,A.\,Schneeloch}
\affiliation{\footnotesize \mbox{Condensed Matter Physics and Materials Science, Brookhaven National Laboratory, Upton, NY, USA}}

\author{G.\,D.\,Gu}
\affiliation{\footnotesize \mbox{Condensed Matter Physics and Materials Science, Brookhaven National Laboratory, Upton, NY, USA}}

\author{F.\,L\'{e}gar\'{e}}
\affiliation{\footnotesize Centre \'{E}nergie Mat\'{e}riaux T\'{e}l\'{e}communications, Institut National de la Recherche Scientifique, Varennes, Qu\'{e}bec J3X 1S2, Canada}

\author{A.\,F.\,Kemper}
\affiliation{\footnotesize \mbox{Department of Physics, North Carolina State University, Raleigh, NC 27695, U.S.A.}}

\author{Ke-Jin\,Zhou}
\affiliation{\footnotesize Diamond Light Source, Harwell Campus, Didcot OX11 0DE, United Kingdom}

\author{V.\,Bisogni}
\affiliation{\footnotesize \mbox{National Synchrotron Light Source II,
Brookhaven National Laboratory, Upton, NY 11973, USA}}

\author{S.\,Blanco-Canosa}
\affiliation{\footnotesize \mbox{Donostia International Physics Center, DIPC, 20018 Donostia-San Sebastian, Basque Country, Spain}}
\affiliation{\footnotesize IKERBASQUE, Basque Foundation for Science, 48013 Bilbao, Spain}

\author{A.\,Frano}
\affiliation{\footnotesize Department of Physics, University of California San Diego, La Jolla, California 92093, USA}
\affiliation{\footnotesize Canadian Institute for Advanced Research, Toronto, ON, M5G 1M1, Canada}

\author{F.\,Boschini}
\affiliation{\footnotesize Centre \'{E}nergie Mat\'{e}riaux T\'{e}l\'{e}communications, Institut National de la Recherche Scientifique, Varennes, Qu\'{e}bec J3X 1S2, Canada}
\affiliation{\footnotesize Quantum Matter Institute, University of British Columbia, Vancouver, BC V6T 1Z4, Canada}

\author{E.\,H.\,da Silva Neto}
\email[Corresponding Author: ]{eduardo.dasilvaneto@yale.edu}
\affiliation{\footnotesize \mbox{Department of Physics, Yale University, New Haven, Connecticut 06520, USA}}
\affiliation{\footnotesize \mbox{Energy Sciences Institute, Yale University, West Haven, Connecticut 06516, USA}}

\maketitle

\large \noindent \textbf{\uppercase{Abstract}}
\normalsize

{In the study of dynamic charge order correlations in the cuprates, most high energy-resolution resonant inelastic x-ray scattering (RIXS) measurements have focused on momenta along the high-symmetry directions of the copper oxide plane. 
However, electron scattering along other in-plane directions should not be neglected as they may contain information relevant, for example, to the origin of charge order correlations or to our understanding of the isotropic scattering responsible for strange metal behavior in cuprates.} We report high-resolution resonant inelastic x-ray scattering (RIXS) experiments that reveal the presence of dynamic electron correlations over the $q_x$-$q_y$ scattering plane in underdoped \bscco~with $T_c=54$\,K. We use the softening of the RIXS-measured bond stretching phonon line as a marker for the presence of charge-order-related dynamic electron correlations. The experiments show that these dynamic correlations exist at energies below approximately $70$\,meV and are centered around a quasi-circular manifold in the $q_x$-$q_y$ scattering plane with radius equal to the magnitude of the charge order wave vector, \qco. 
{We also demonstrate how this phonon-tracking procedure provides the necessary experimental precision to rule out fluctuations of short-range directional charge order (i.e. centered around $[q_x=\pm q_{CO}, q_y=0]$ and $[q_x=0, q_y=\pm q_{CO}]$) as the origin of the observed correlations.}

\vspace{5mm}

\large \noindent \textbf{\uppercase{Introduction}}
\normalsize

Dynamic fluctuations from periodic charge order (CO) pervade the phase diagram of cuprate superconductors, perhaps even more than superconductivity itself \cite{Arpaia_review}. The detection of these fluctuations over energy and momentum was enabled by several recent advances in the energy resolution of resonant inelastic x-ray scattering (RIXS) instruments operating in the soft x-ray regime.
In the case of \ybco, Cu-L$_3$ RIXS detects dynamic correlations at the charge order wavevector, \qco, with a characteristic energy scale of approximately $20$\,meV \cite{Arpaia_Science}. 
It has been proposed that these low-energy short-range dynamic charge order correlations are a key ingredient to the strange metal behavior \cite{seibold2021strange,caprara2022dissipation} {characterized by linear-in-temperature resistivity \cite{T_linear_cuprates_1987, T_linear_BSCO_1989}. On one hand, this temperature behavior is often associated with an isotropic scattering rate that depends only on temperature in units of energy and Planck's constant (\textit{i.e.}\,$\propto k_B T / \hbar$, sometimes called the Planckian regime) \cite{Varma_MFL, Varma_critical, Patel_PRX, Patel_PRL, Phillips_Science_Review}, as supported by recent angle-dependent magnetoresistance measurements of \lnsco~\cite{grissonnanche2021linear}. On the other hand, combined transport and RIXS studies have recently shown an unexpected link between linear-in-temperature resistivity and charge order in \ybco~\cite{Wahlberg_Science_2021, LeTacon_Comment}.} 
Combined, these latest results suggest that fluctuations of the charge order should somehow result in an effective isotropic scattering. Still, high-resolution RIXS experiments have largely focused on the fluctuations along the high-symmetry crystallographic directions only, leaving 
the full structure of electron correlations within the copper oxide plane unknown. 


Recently, in \bscco~(Bi-2212), RIXS measurements 
found the existence of a quasi-circular pattern in the $q_x$-$q_y$ plane at finite energies and with the same wave vector magnitude as that of the observed static charge order peak at $\mathbf{q} = [q_x=\pm q_{CO}, q_y=0]$ and $[q_x=0, q_y=\pm q_{CO}]$ -- \textit{i.e.} dynamic correlations with charge order wavelength along all direction in the CuO$_2$ plane \cite{boschini2021dynamic}.
Although the medium energy-resolution of those measurements ($\Delta E \approx0.8$\,eV) precluded a more precise determination of their energy profile, {the results suggested} that these quasi-circular dynamic correlations (QCDCs) appear broad over the mid-infrared ranges (defined approximately as $100$ to $900$ meV).
This scattering manifold, which may result from combined short- and long-range Coulomb interactions \cite{boschini2021dynamic, Yamase_PRB_2021, Bejas_ringlike_2022}, would provide a large variety of wave vectors for connecting all points of the Fermi surface (\textit{i.e.} an effective isotropic scattering). However, it is not yet experimentally known if this manifold extends to electron scattering at lower energies, in the quasi-elastic regime.
To experimentally investigate this scenario we used high energy-resolution ($\approx37$\,meV) Cu-L$_3$ RIXS $q_x$-$q_y$ mapping of the electronic correlations in Bi-2212. Using the softening of the bond stretching (BS) phonon in RIXS as a marker of charge order correlations, our measurements reveal the presence of low-energy quasi-circular dynamic electronic correlations with $|\mathbf{q}| \approx q_{CO}$.

\vspace{5mm}

\large \noindent \textbf{\uppercase{Results}}
\normalsize

\vspace{2mm}

\noindent \textbf{High-resolution RIXS mapping of dynamic correlations in the $q_x$-$q_y$ plane}

\vspace{2mm}

\begin{figure}
    \centering
    \includegraphics[width=\textwidth]{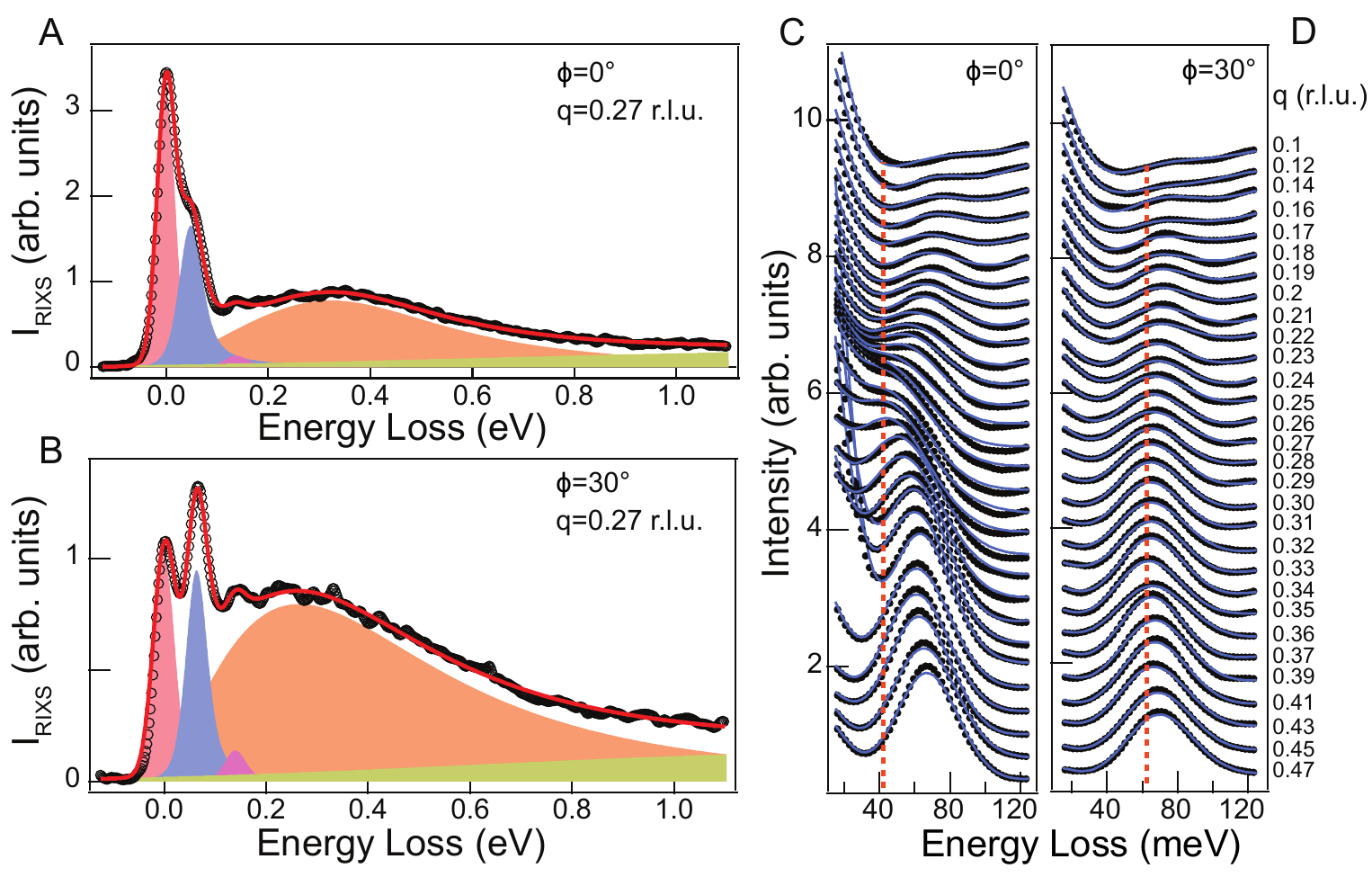}
    \caption{\textbf{RIXS spectra and fitting.} (\textbf{A} and \textbf{B}) Examples of spectra at $q = 0.27$\,r.l.u.~for $\phi = 0^\circ$ and $30^\circ$, respectively (open circles). The red lines are fits to the spectra, composed of a quasi-elastic peak (pink), a BS phonon peak at $\approx 70$\,meV (blue), a peak at $\approx 135$\,meV (likely from a two-phonon process) (purple), a broad paramagnon (orange) and a broad background feature of unknown origin (brown). (\textbf{C} and \textbf{D}) RIXS measured BS phonon peak for various values of $q$ measured for $\phi = 0^\circ$ and $30^\circ$, respectively (black circles). The blue lines are the fits to the spectra. The vertical orange dashed lines, indicating the lowest phonon peak position at each $\phi$, are shown to help the reader observe the phonon dispersions in the raw data. }
    \label{F1}
\end{figure}

We performed measurements at $\phi = 0^\circ, 25^\circ, 30^\circ, 35^\circ, 45^\circ$, where $\phi$ is defined as the azimuthal angle from the $q_x$ axis. For each $\phi$, we acquired RIXS spectra at different values of in-plane momentum-transfer $q = |\mathbf{q}|$ by varying the incident angle on the sample. Throughout the paper, values of $q$ are reported in reciprocal lattice units (r.l.u.), where one r.l.u.~is defined as $2\pi/a$ and $a = 3.82$\,\AA~(the lattice constant along $\phi = 0^\circ$). In Fig.\,\ref{F1} (A and B), we show representative spectra obtained at $q$ near $q_{CO}$ for $\phi = 0^\circ$ and $30^\circ$, and energies below $1.1$\,eV.  In these two cases, the minimal model that fits the data includes five contributions: a quasi-elastic peak, a bond-stretching phonon peak at $\approx 70$\,meV, a peak at $\approx 135$\,meV (likely from a two-phonon process), a broad paramagnon and a broad background feature of unknown origin. A similar assessment can be made regarding all other high-resolution spectra acquired in this work. In this type of fitting analysis, the QCDCs are not explicitly accounted for and it is generally difficult to disentangle overlapping contributions to the RIXS spectra using a fitting model with so many parameters, thus precluding the extraction of the exact spectral profile of the QCDCs with any reasonable confidence. Still, we note that this high-resolution data is consistent with the previously reported medium-resolution data \cite{boschini2021dynamic}, which can be verified by integration of the high-resolution data (see supplementary materials, Fig.\,S7).

\begin{figure}
    \centering
    \includegraphics[width=0.66\textwidth]{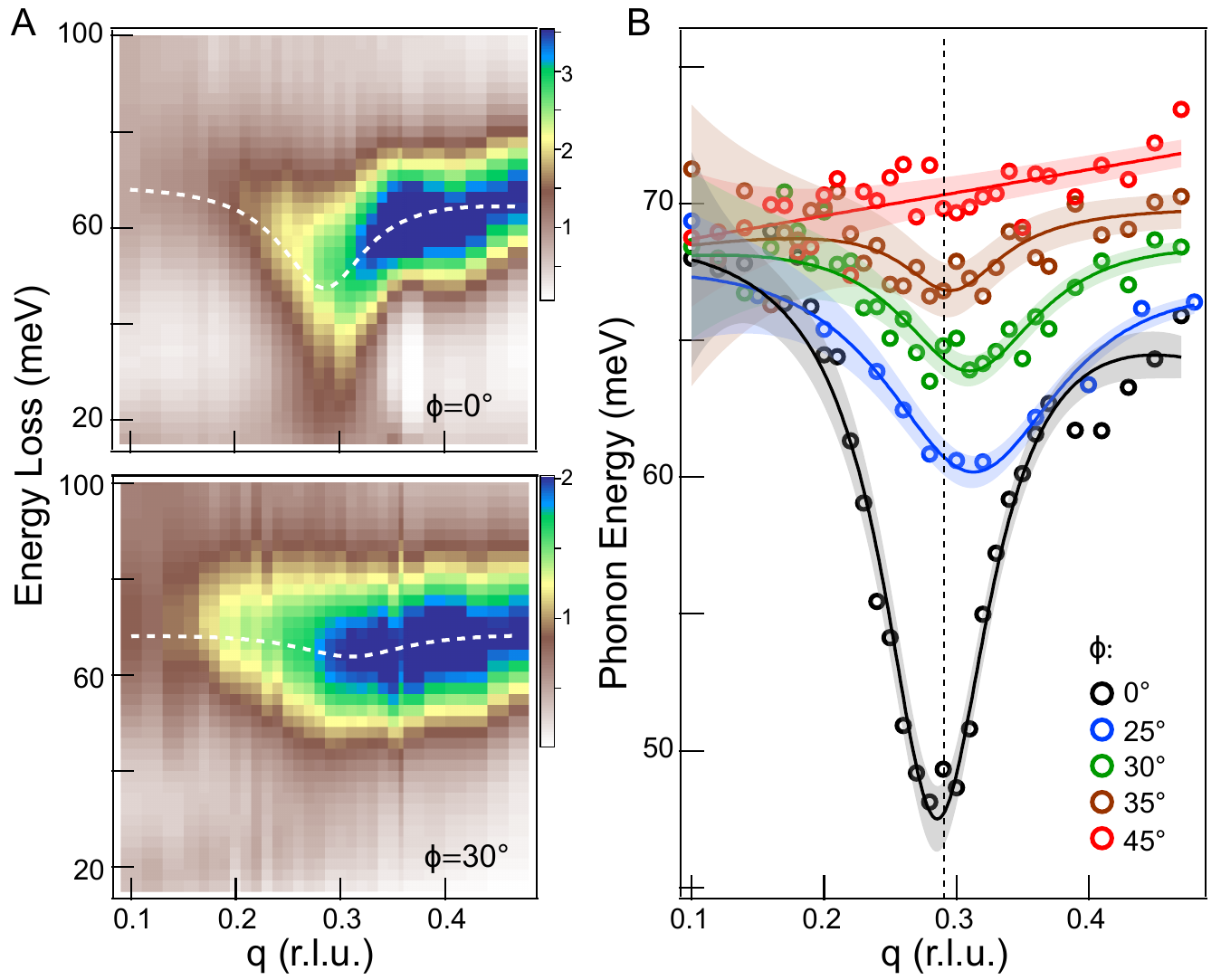}
    \caption{\textbf{Location of low-energy dynamic correlations extracted from the phonon dispersion.} (\textbf{A}) Energy-momentum structure of the excitations at $\phi = 0^\circ$ and $30^\circ$ after subtraction of the elastic line. The image is constructed from RIXS spectra deconvoluted from the energy resolution. (\textbf{B}) Location of the phonon peak obtained by fitting the RIXS spectra deconvoluted from energy resolution for different $\phi$ (see Materials and Methods and also Supplementary Materials, Fig.\,S3). The solid lines are obtained by fitting the $q$-dependence of the phonon peak (circles) with a negative Lorentzian function plus a linear background. The shaded regions around the solid lines are generated from the 95\% confidence interval obtained for the various fits to the spectra (see Materials and Methods for details). The solid lines for $\phi = 0^\circ$ and $30^\circ$ in (B) appear as dashed white lines in \textbf(A).}
    \label{F2}
\end{figure}

It is likely that the spectral intensity of QCDCs in Bi-2212 is so dilute over energy as to preclude the extraction of their spectral structure amidst stronger paramagnon and phonon signals. Still, here we develop a different method to detect QCDCs at lower energies, by tracking the evolution of the bond-stretching (BS) phonon over the $q_x$-$q_y$ plane. This method is based on the phenomenology revealed by several recent RIXS measurements of the cuprates along $q_x$ and $q_y$, which indicate an apparent softening of the BS phonon peak at the momentum location of the static CO peak \cite{Chaix2017,Lin_PRL_2020,Peng_PRL_2020,Li_PNAS_2020,Wang_SciAdv_2021,lee2021melting}. In the case of Bi-2212, it has been proposed that the apparent softening of the BS phonon in RIXS is due to an interplay between low-energy fluctuations of the charge order and BS phonons that results in a Fano-like interference \cite{Chaix2017,Li_PNAS_2020, lee2021melting, Lu_PRB_2022_phonons_2212}. {Another possibility is that the apparent softening is simply the result of the phonon peak and a low-energy charge order peak overlapping, as recently suggested by measurements of both \ybco~and Bi-2212 \cite{Arpaia_arXiv}.} 
In either interpretation the location of the phonon softening can be used as a marker for low-energy charge order correlations.

Figure \ref{F1} (C and D) shows the spectra acquired as a function of $q$ for $\phi = 0^\circ$ and $30^\circ$, respectively, focusing on the region of the BS phonon. At $\phi = 0^\circ$, it is clear that the phonon peak position softens to its lowest energy value at $q = q_{CO} \approx 0.29$\,r.l.u.~(Fig.\,\ref{F1}C). Careful observation of the spectra taken along $\phi = 30^\circ$ shows a similar softening effect with the lowest phonon energy position occuring for $q \approx q_{CO} $~(Fig.\,\ref{F1}D). 
Figure \ref{F2}A shows the mapping of the BS phonon mode at $\phi = 0^\circ$ and $30^\circ$ obtained after subtraction of the fitted elastic line, once again showing the softening of the RIXS phonon even at $\phi = 30^\circ$. To precisely determine the locations of the softening in the $q_x$-$q_y$ plane, we fit the spectra to extract the dispersion of the BS phonon for each $\phi$ (Fig.\,\ref{F2}B). {We observe a softening of the RIXS measured phonon line for all $\phi$, except for $\phi = 45^\circ$. Remarkably all observed softening occurs at a value of $q \approx q_{CO}$, precisely as expected for QCDCs at low energies.}

\begin{figure}
    \centering
    \includegraphics[width=1\textwidth]{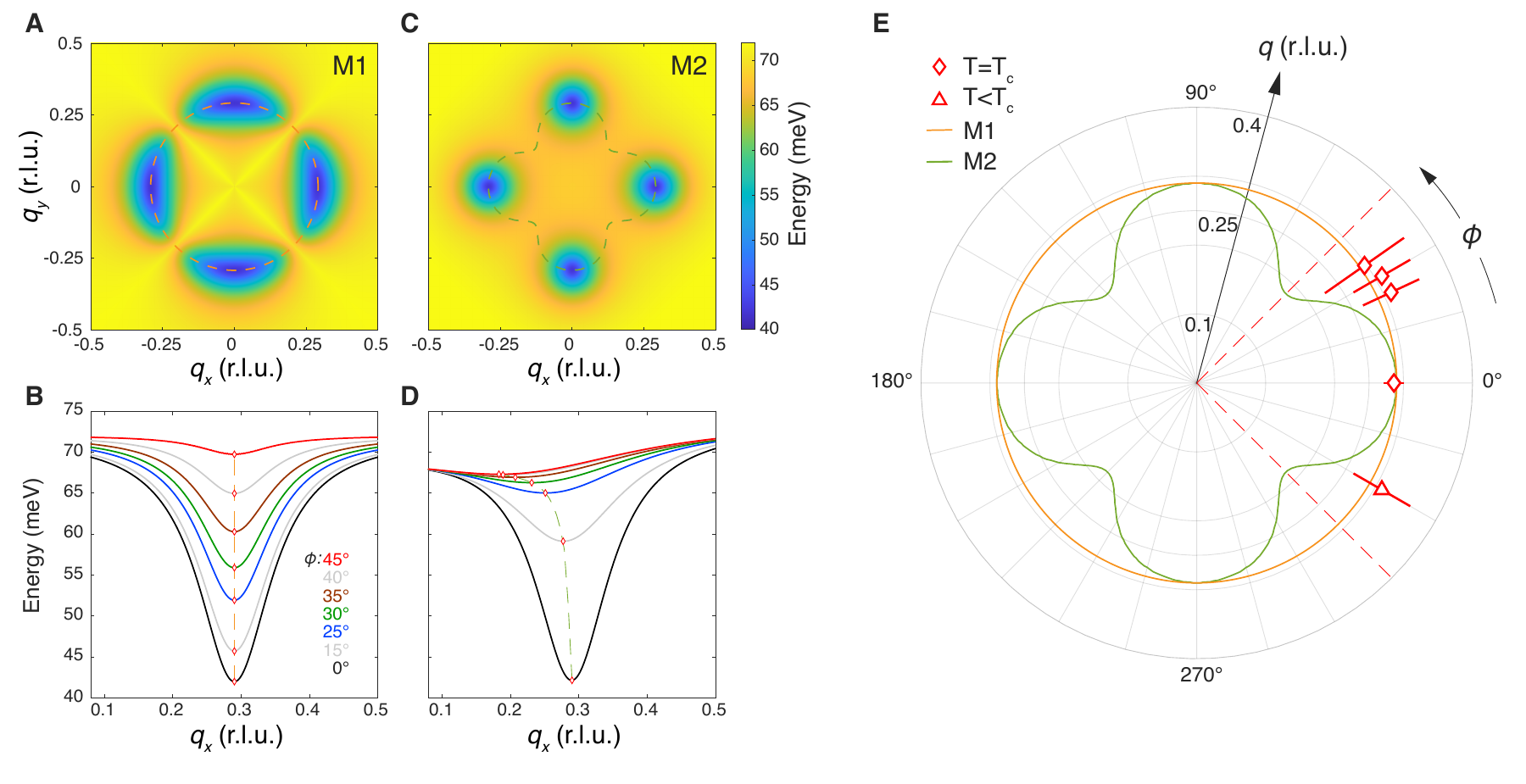}
    \caption{\textbf{Models of phonon softening for QCDCs and directional order.} (\textbf{A} and \textbf{C}) Phonon dispersion for M1 and M2, as described in the text. (\textbf{B} and \textbf{D}) Momentum $q$ cuts of the phonon dispersion at different $\phi$ for the simulated data in (A) and (C) respectively. The dashed orange and green lines in (A-D) identify the location of the phonon softening in the $q_x$-$q_y$ plane. (\textbf{E}) Polar plot contrasting M1, M2 models (orange and green solid lines) and the experimental data (red symbols). The error bars in (E) are obtained from the fits to the phonon dispersion in Fig.\,\ref{F2}B. See Materials and Methods for more details.}
    \label{F3}
\end{figure}

\vspace{2mm}

\noindent \textbf{Discriminating QCDCs from short-range directional order}

\vspace{2mm}

Dynamic correlations emanating from short range order are bound to be broad in $\mathbf{q}$. It is therefore reasonable to ask whether the measured $q_x$-$q_y$ profile of the BS phonon could simply be the result of diffuse scattering from short-range directional order. The fundamental difference between QCDCs and short-range directional order is that the former forms a manifold of dynamic correlations centered at $q = q_{CO}$ {(similar to Brazovskii-type fluctuations \cite{boschini2021dynamic, Brazovskii})}, while the the latter results in dynamic correlations around $\mathbf{q} = [q_x=\pm q_{CO}, q_y=0]$ and $\mathbf{q} = [q_x=0, q_y=\pm q_{CO}]$ (more details on M1 and M2 are provided in the Materials and Methods section). To contrast these scenarios we consider two simple toy models. In both cases we start with a flat $|\mathbf{q}|$-independent phonon mode at $72$\,meV, which is a reasonable approximation given the small dispersion of the BS phonon in the absence of charge order \cite{Giacomo_PRR_phonon_flat, Arpaia_arXiv}. 
{In the first model (M1) we construct the QCDCs scenario, where the $q$-cuts for various $\phi$ always have a minimum located at $q = q_{CO}$, Fig.\,\ref{F3}B.} 
In the second model (M2) we consider the case where dynamic charge order correlations emerge isotropically from static peaks at $[q_x=\pm q_{CO}, q_y=0]$ and $[q_x=0, q_y=\pm q_{CO}]$. The corresponding phonon profile is shown in Fig.\,\ref{F3} (C and D). 
{To roughly emulate the data we also introduce a $\phi$-dependent phonon minimum in M1, which increases from $\phi = 0^\circ$ to $45^\circ$, Fig.\,\ref{F3}A.} 
{However, note that the magnitude of the softening depends on the $\phi$ structure of the electron-phonon coupling, which is not known or necessary for discerning the two scenarios.}
The $q$-cuts show a qualitatively similar behavior in both models: a clear phonon softening at $\phi = 0$ that continues to exist even as $\phi$ approaches $45^\circ$. However, in M2 the $q$ location of the phonon minima clearly decreases with increasing $\phi$ from $0^\circ$ to $45^\circ$. This comparison explains our selection of $\phi$ values for these studies: the experimental ability to differentiate between M1 and M2 is largest in the $\phi = 25^\circ$ to $45^\circ$ range. 
The polar plot in Fig.\,\ref{F3}E summarizes the analysis, comparing the q-location of the minima for both models to the minima obtained from experiments (red markers). Within the error bars, the RIXS measurements are consistent with M1 and rule out M2, indicating the quasi-circular nature of the low-energy correlations associated with the charge order.


\vspace{5mm}

\large \noindent \textbf{\uppercase{Discussion}}
\normalsize

\vspace{2mm}

The experiments presented here provide evidence for the existence of quasi-circular dynamic correlations at low energies in underdoped Bi-2212, which could be a key ingredient for models that connect charge order to an effective isotropic scattering. Long-range translational symmetry breaking cannot be responsible for this isotropy due to the characteristic length scale and directionality of the ordered state. Although short-range electron correlations from directional order, occupying a much larger region of momentum space, could in principle emulate isotropic scattering \cite{Abanov_Spin_Fermion, seibold2021strange, caprara2022dissipation}, the QCDCs revealed by our experiments offer a different scenario. {Extending not only around the static charge order wave vectors but also in the azimuthal direction, QCDCs might be a more viable platform for isotropic scattering.}
To fully understand the impact of QCDCs to electronic properties of the cuprates, one requires knowledge of the energy structure of these correlations. Although this might still be beyond the current experimental capabilities, our experiments provide some constraints to the low-energy structure of the QCDCs. 
{In particular, for the $\phi$ values where a softening is detected, QCDCs must exist below $\approx70$\,meV (\textit{i.e.} the approximate energy of the bare phonon). 
Unfortunately the amount of energy softening at $q=q_{CO}$ by itself, without knowledge of the $\phi$-dependence of the electron-phonon interaction, does not provide more information about the energy structure of the QCDCs. 
Therefore, it remains possible that QCDCs at $\phi = 45^\circ$ exist below $70$\,meV but do not significantly interact with the BS phonon.}


The quasi-circular shape of the low-energy correlations is similar to the shape obtained from the analysis of higher energy correlations (Ref.\,\cite{boschini2021dynamic} and Supplementary Materials, Fig.\,S6). This similarity raises the possibility that the QCDCs exist up to much higher energies of order of $1$\,eV. As discussed in Ref.\,\cite{boschini2021dynamic}, the quasi-circular correlations cannot be explained by an instability of the Fermi surface. Instead, it was proposed that the location of the dynamic CO correlations in $\mathbf{q}$-space is determined by the minima of the effective Coulomb interaction, which becomes non-monotonic in $q$ due to the inclusion of a long-range Coulomb interaction. However, this non-monotonic Coulomb interaction by itself failed to capture the intensity anisotropy observed at $q \approx q_{CO}$.  Likewise, here the same proposed Coulomb interaction could also explain the most salient feature of our data, namely the quasi-circular shape of the low-energy correlations. 
Recently, a more complete theoretical description based on a $t$-$J$ model with long-range Coulomb interaction shows the presence of ring-like charge correlations with the correct intensity anisotropy \cite{Bejas_ringlike_2022}.
The results presented here can serve as a guide for future theoretical investigations that also account for the apparent decrease of the phonon softening from $\phi = 0^\circ$ to $45^\circ$.

Beyond the fact that both the energy-integrated correlations \cite{boschini2021dynamic} and the low-energy dynamic correlations appear to occupy the same quasi-circular scattering manifold, the current RIXS measurements do not provide further experimental evidence to connect these two phenomena. Such additional evidence may come from polarimetric RIXS experiments that are able to decompose charge and spin excitations in the mid-infrared range, as it has been done for electron-doped cuprates \cite{daSilvaNeto_PRB_2018}. Compared to the energy-integration procedure, the phonon tracking method provides larger precision for mapping  CO correlations in the $q_x$-$q_y$ plane, since the large integration ranges required for the former result in very broad features in $\mathbf{q}$-space. 
Indeed, we have already performed medium-resolution RIXS measurements that detect the presence of similar quasi-circular scattering manifolds in the energy-integrated spectrum of optimally and overdoped samples, but the investigation of their doping dependence is hindered by the large experimental uncertainty associated with the integration method (see Supplementary Materials, Fig.\,S6). 
\textcolor{black}{Instead, our new procedure to track QCDCs using measurements of the RIXS BS phonon goes beyond demonstrating the existence of QCDCs in underdoped Bi-2212 at low energies. It is also a new methodology that can be used to detect QCDCs in other cuprates and understand related phenomena such as the electron-doped cuprates which also show a quasi-circular scattering~\cite{Kang2019}.}
Finally, the application of this new method to multiple cuprate families at different dopings and/or temperatures will help unveil whether and how QCDCs and the strange metal are related.



\vspace{5mm}

\large \noindent \textbf{\uppercase{Materials and Methods}}
\normalsize

\vspace{4mm}

\noindent \textbf{RIXS experiments}\\
High-resolution RIXS experiments were performed at the I21 beamline \cite{Zhou:rv5159} at Diamond Light Source, United Kingdom, and at the 2-ID beamline at the National Synchrotron Light Source II, Brookhaven National Laboratory, USA. The orientation of the crystal axes of the underdoped \bscco~samples with $T_c = 54$\,K was obtained by x-ray diffraction prior to the RIXS experiment. The samples were cleaved in air just moments before inserting them into the ultra-high-vacuum chambers. For experiments at I21, the crystal was aligned to the scattering geometry \textit{in situ} from measurements of the {002} Bragg reflection and the b-axis superstructure peak. The scattering angle was fixed at $154^\circ$ (I21) and $153^\circ$ (2-ID). The incoming light was set to vertical polarization ($\sigma$ geometry) at the Cu-$L_3$ edge ($\approx 931.5$\,eV). The combined energy resolution (FWHM) was about $37$\,meV (I21) and $40$\,meV (2-ID), with small variations ($\pm 3$\,meV) over the course of multiple days. In both cases the energy resolution was relaxed in a trade-off for intensity. The projection of the momentum transfer, $q$, $q_x$-$q_y$ plane was obtained by varying the incident angle on the sample ($\theta$). All the measurements were performed at $T=54\,K$, which is the superconducting transition temperature for this sample, except for one measurement performed at {$T=25$\,K below $T_c$ (Fig.\,\ref{F3})E} and one measurement at $300$\,K (Supplementary Materials, Figs.\,S2 and S5).

\noindent \textbf{Analysis of RIXS spectra}\\
To ensure the robustness of the extraction of phonon dispersion from the RIXS spectra we analyzed the data using multiple methods. Although the overall RIXS cross-section may depend on $\phi$, we did not perform any normalization or intensity correction procedure to the spectra since the energy location of the phonon does not depend on the overall intensity. A comparison between the results for different methods is available in the supplementary materials, Fig.\,S4.\\
\textbf{Method 1:} In an effort to maintain an agnostic approach and to not assume particular functional forms of the different contributions to the RIXS spectra, we extracted the dispersion by simply tracking the energy positions of the phonon peak maximum in the RIXS spectra deconvoluted from the energy resolution. See below for details of the deconvolution procedure. \\ 
\textbf{Method 2:} The phonon dispersion shown in Fig.\,\ref{F2}B was extracted by fitting the deconvoluted RIXS spectra (see Fig.\,\ref{F2}A and supplementary materials, Fig.\,S3) in the [-30,130]\,meV range to a double Gaussian function plus a second order polynomial background, keeping all parameters free. The shaded regions around the solid lines in Fig.\,\ref{F2}B were generated by fitting the $95$\% confidence intervals (obtained from the fits) to a polynomial function of $q$.\\
\textbf{Method 3:} Following previous works \cite{lee2021melting,Wang_paramagnons_2022}, the raw RIXS spectra were fit to a five component model that includes a Gaussian (elastic peak of amplitude $A_{\text{el}}$, position $\omega_{\text{el}}$ and width $w_{\text{el}}$), two anti-Lorentzians (phonon and double-phonon peaks of different amplitude $A_{\text{i}}$ and position $\omega_{\text{i}}$, and sharing width $w_{\text{ph}}$ and Fano parameter width $q_{\text{F}}$ -- note that $i$=1,2 indicates the first and second phonon, respectively), a damped harmonic oscillator lineshape (paramagnon of amplitude $A_{\text{pm}}$, position $\omega_{\text{pm}}$ and damping parameter $\gamma_{\text{pm}}$), and an error function (smooth background described by an error function with amplitude amplitude $A_{\text{BG}}$, position $\omega_{\text{BG}}$ and width $w_{\text{BG}}$):
\begin{equation}
\begin{split}
    f(\omega)= A_{\text{el}} e^{-\frac{(\omega-\omega_{\text{el}})^2}{w_{\text{el}}^2}} + \sum_{\text{i=1,2}} A_{\text{i}}\frac{2(\omega-\omega_{\text{i}})/w_{\text{ph}}+q_{\text{F}}}{[2(\omega-\omega_{\text{i}})/w_{\text{ph}}]^2+1} + \\ +A_{\text{pm}}\frac{\gamma_{\text{pm}}\omega}{(\omega^2-\omega_{\text{pm}}^2)^2+4\gamma_{\text{pm}}^2\omega^2} + A_{\text{BG}}\Big[\textrm{erf}\Big(\frac{\omega-\omega_{\text{BG}}}{w_{\text{BG}}}\Big)+1\Big]
\end{split}
\end{equation}
The fitting model is convoluted with the RIXS energy resolution ($\sim$37\,meV). From this analysis we extracted the phonon dispersion for each $\phi$ that quantitatively matches the phonon dispersion shown in Fig.\,\ref{F2}B. All parameters are kept free, except for $\omega_{bg}$, which is constrained within a range of $[0.2, 0.6]$\,eV. 

\noindent \textbf{Fitting the phonon dispersion}\\
To obtain a phenomenological form to the dispersions in Fig.\,\ref{F2}B, the extracted peak locations as a function of $q$ were fit to a linear background plus a negative Lorentzian function. From this fit we obtain the $q$ location of the softening (red markers in Fig.\,\ref{F3}E). To obtain the error bars in Fig.\,\ref{F3}E, we follow a conservative approach by taking the average of the two $q$-intercepts of the fitted curve at $E_{min} + 2$\,meV, where $E_{min}$ is the lowest energy of the dispersion and $\pm2$\,meV is the typical amount of scatter observed in the data. For $\phi = 45^\circ$ the data is fit to a line.

\noindent\textbf{Deconvolution procedure}\\
We employed the Lucy-Richardson deconvolution procedure \cite{yang_Deconv_2008} to deconvolve the energy resolution ($\sim$37\,meV) from the RIXS spectra (deconvoluted curves for all azimuths are displayed in the supplementary materials, Fig.\,S3). The number of iterations and region of interest of the deconvolution procedure were optimized by ensuring that the convolution of the deconvoluted curves reproduced the raw data. 

\noindent \textbf{Model simulations}\\
The toy models M1 and M2 are purely phenomenological. For M1, the phonon dispersion was modeled as:
\begin{equation}
    E = E_0 - \xi(\phi) \frac{\Delta}{(\frac{q-q_0}{\Gamma})^2 + 1}
\end{equation}
where $E_0 = 72$\,meV, $\Delta = 30$\,meV, $\Gamma = 0.065$\,r.l.u., $q_0 = 0.29$\,r.l.u.~and $\xi(\phi) = (|\cos(2\phi)| + 0.08)/(1.08)$. 
For M2, the phonon dispersion was modeled as: 
\begin{equation}
    E = E_0 - \sum_{i=1}^{4} \frac{\Delta}{(\frac{\mathbf{q}-\mathbf{q}_i}{\Gamma})^2 + 1}
\end{equation}
where $E_0 = 74$\,meV, $\Delta = 30$\,meV, $\Gamma = 0.065$\,r.l.u. and $\mathbf{q}_i$ are the four peaks located at $[q_x=\pm q_{CO}, q_y=0]$ and $[q_x=0, q_y=\pm q_{CO}]$ ($q_{CO} = 0.29$\,r.l.u.)

\vspace{2mm}

\begin{acknowledgments}

We acknowledge the Diamond Light Source for time on beamline I21-RIXS under proposals MM28523 and MM30146. 
This research used resources of the Advanced Light Source, a DOE Office of Science User Facility under contract no. DE-AC02-05CH11231.
This research used beamline 2-ID of the National Synchrotron Light Source II, a U.S. Department of Energy (DOE) Office of Science User Facility operated for the DOE Office of Science by Brookhaven National Laboratory under Contract No. DE-SC0012704. 
We especially acknowledge the incredible work done by the beamline staffs at I-21, 2-ID and at 8.0.1 qRIXS, to allow many of these experiments to be performed remotely during the COVID pandemic. This work was supported by the Alfred P. Sloan Fellowship (E.H.d.S.N.). E.H.d.S.N acknowledges support by the National Science Foundation under Grant No.\,2034345. A.F.K. was supported by the National Science Foundation under grant no. DMR-1752713. F.B. acknowledges support from the Fonds de recherche du Québec – Nature et technologies (FRQNT) and the Natural Sciences and Engineering Research Council of Canada (NSERC). A.F. was supported by the Research Corporation for Science Advancement via the Cottrell Scholar Award (27551) and the CIFAR Azrieli Global Scholars program. This material is based upon work supported by the National Science Foundation under Grant No. DMR-2145080. The synthesis work at Brookhaven National Laboratory was supported by the US Department of Energy, oﬃce of Basic Energy Sciences, contract no. DOE-SC0012704. 
\end{acknowledgments}

\newpage

\section*{Supplementary Materials}

\renewcommand{\thefigure}{S\arabic{figure}}
\setcounter{figure}{0}    

\begin{figure}[H]
    \centering
    \includegraphics[width=0.8\textwidth]{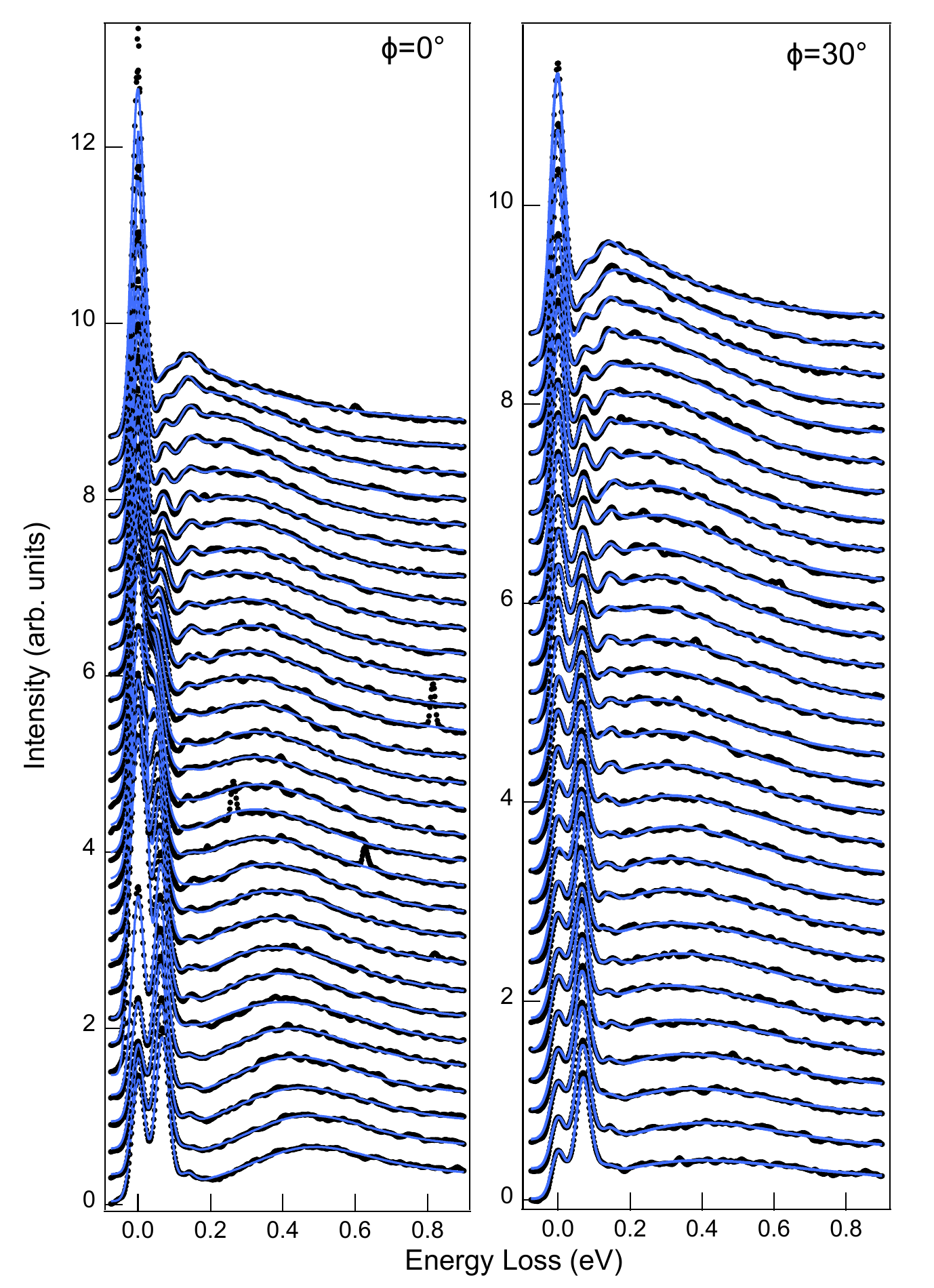}
    \caption{\textbf{Fits to the spectra in Fig.\,2} The two panels show the fit of the RIXS spectra over a wide energy range using Method (3) as detailed in the Materials and Methods Section of the main text.  }
    \label{FSM1}
\end{figure}

\newpage

\begin{figure}[H]
    \centering
    \includegraphics[width=\textwidth]{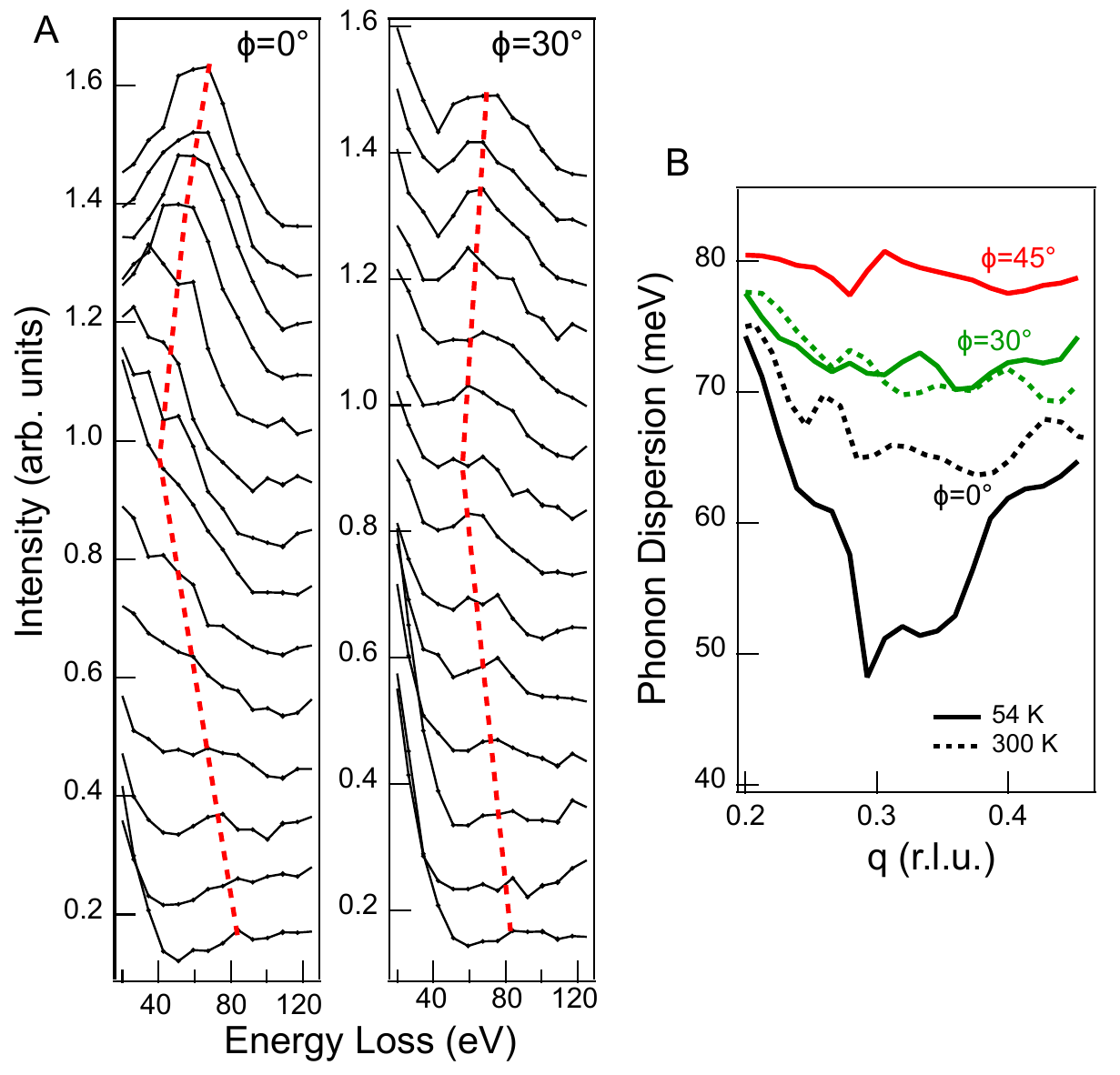}
    \caption{\textbf{Data obtained at the 2-ID beamline at NSLS-II.} (\textbf{A}) RIXS spectra measured as a function of $q$ for different $\phi$ at $54$\,K. The dashed line is a guide to the eye highlighting the phonon softening, visible at both $\phi=0^\circ$ and $\phi=30^\circ$ already from the raw data without need of doing any deconvolution. (\textbf{B}) Phonon dispersion obtained by using Method (2) as detailed in the Materials and Methods Section of the main text. }
    \label{FSM2}
\end{figure}

\newpage

\begin{figure}[H]
    \centering
    \includegraphics[width=\textwidth]{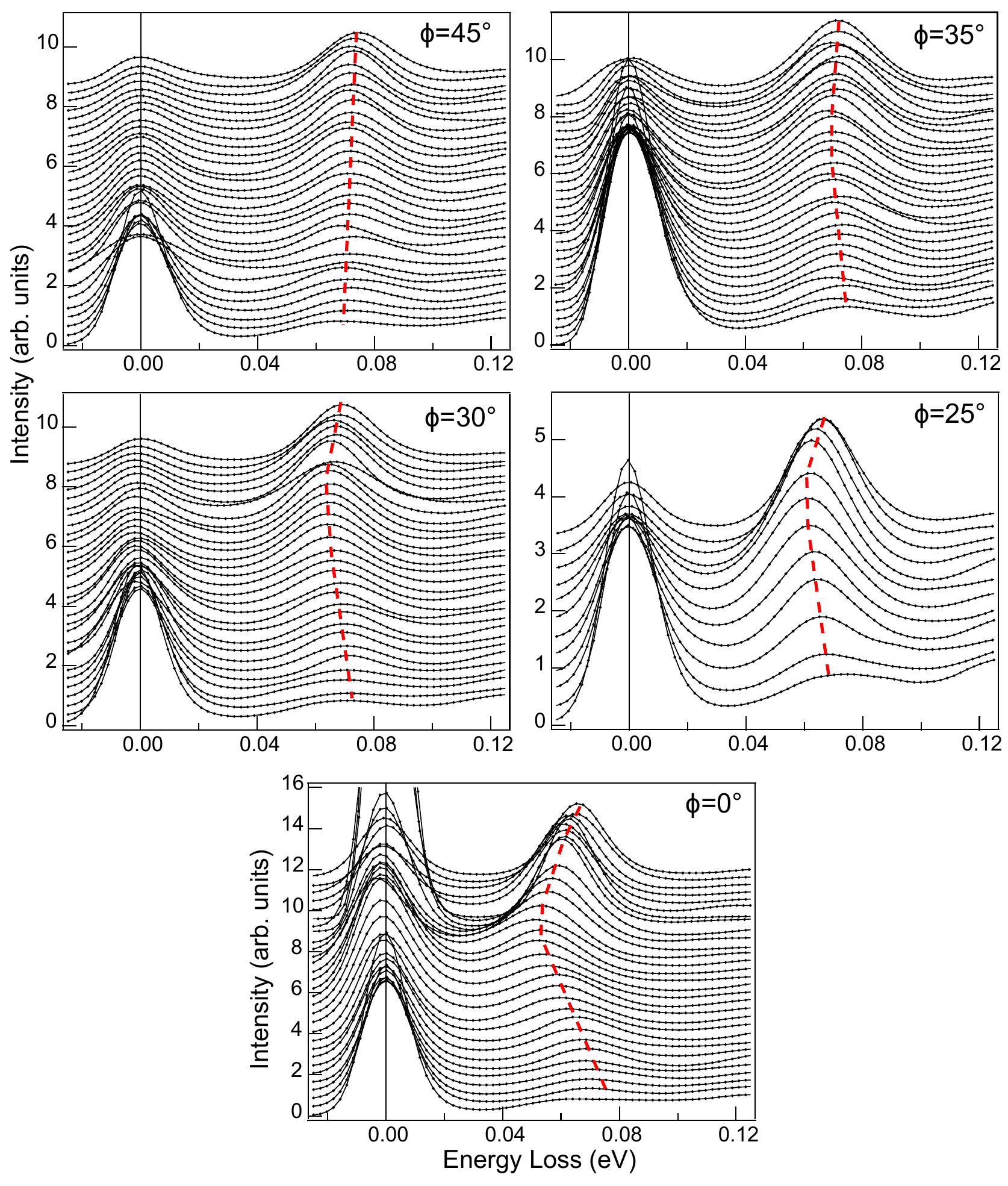}
    \caption{\textbf{Waterfall plot of the RIXS spectra deconvoluted for energy resolution (37\,meV) for different $\phi$.} {The softening of the BS phonon is clearly visible for any $\phi$ (except $\phi$=45$^o$) by simple visual inspection of the deconvoluted curves (red dash lines are guides to the eye). } }
    \label{FSM3}
\end{figure}

\newpage

\begin{figure}[H]
    \centering
    \includegraphics[width=\textwidth]{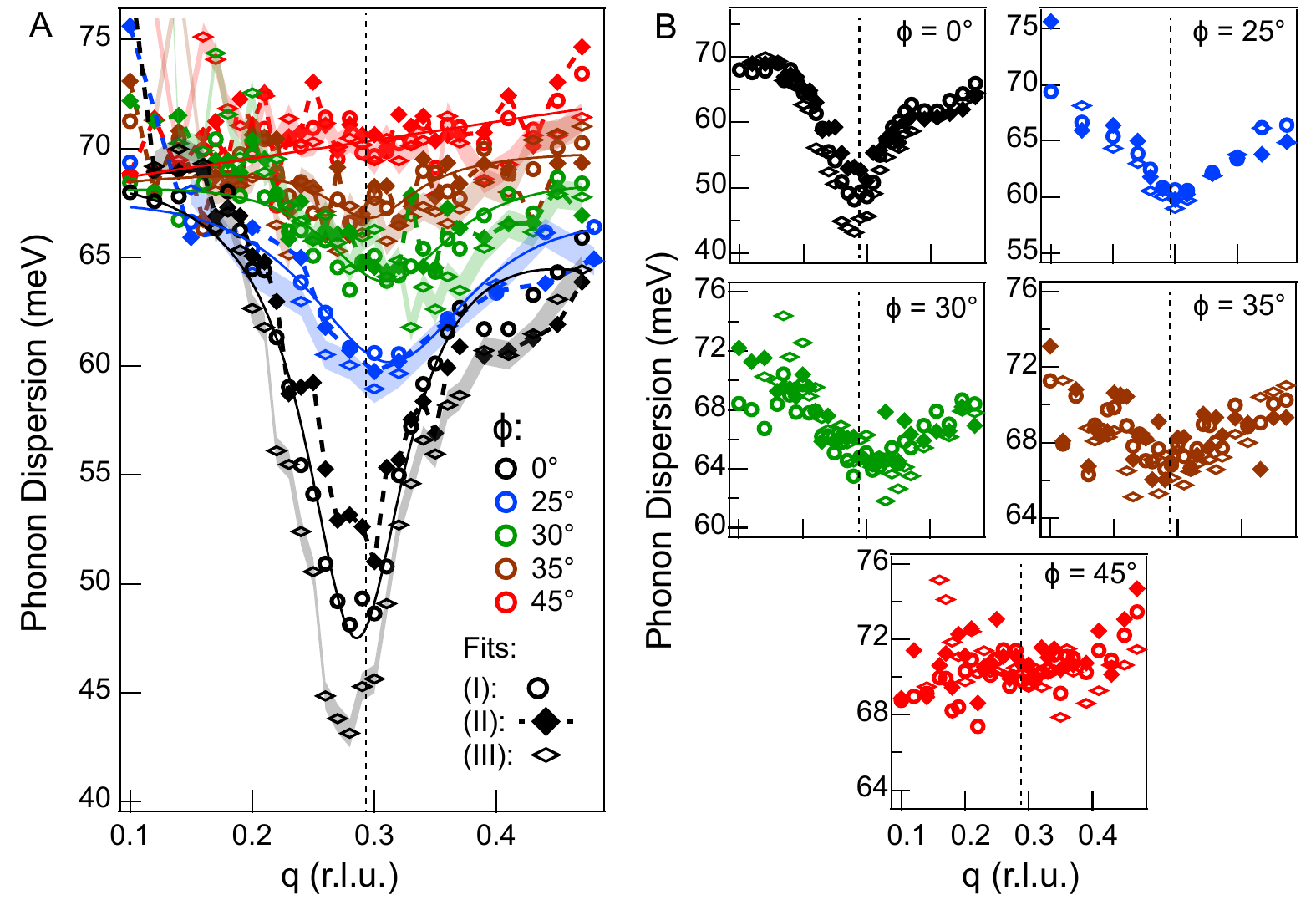}
    \caption{\textbf{Comparison between three different methods for the extraction of the phonon dispersion.}  Methods (1), (2) and (3) are detailed in the Materials and Methods section of the main text. }
    \label{FSM4}
\end{figure}

\newpage

\begin{figure}[H]
    \centering
    \includegraphics[width=0.75\textwidth]{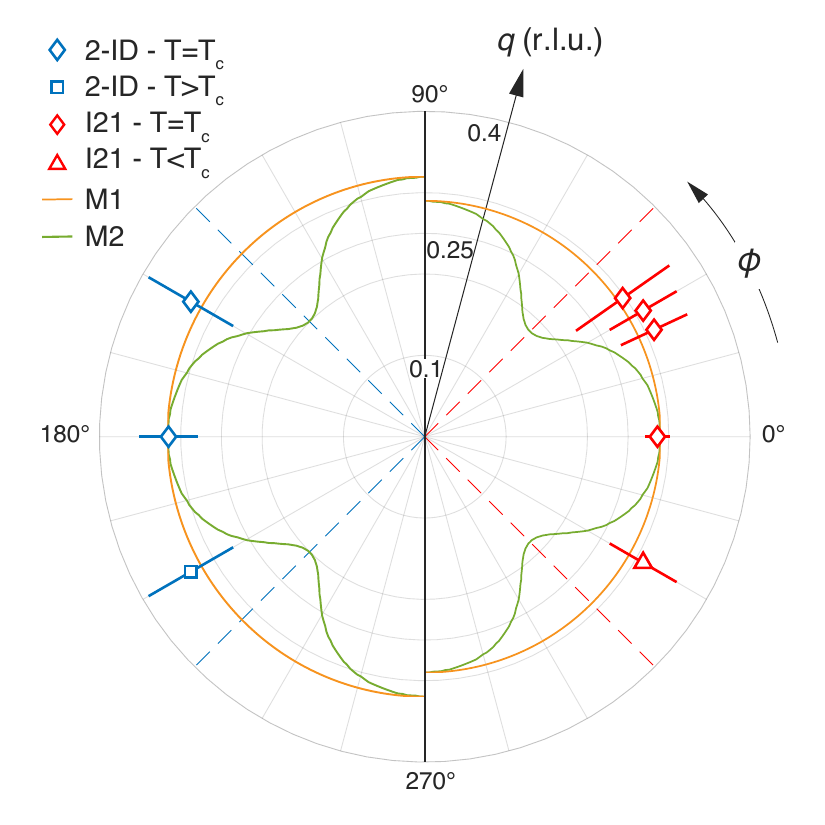}
    \caption{\textbf{Comparison of models to experiments at 2-ID and I21.} Polar plot contrasting M1, M2 models (orange and green solid lines) and the experimental data (blue and red symbols). The error bars are obtained from the fits to the phonon dispersion as described in the Materials and Methods section. On the left side, the model was adjusted for a higher value of $q_{\text{CO}}$ for comparison with the data obtained at 2-ID. The data at 2-ID is consistent with M1 and not with M2. In the main text only the data from I21 is shown because for those experiments the sample crystal axes could be aligned \textit{in situ} from structural diffraction peaks by using the photodiode detector in that chamber. See Materials and Methods section for details.}
    \label{FSM5}
\end{figure}

\newpage

\begin{figure}[H]
    \centering
    \includegraphics[width=0.8\textwidth]{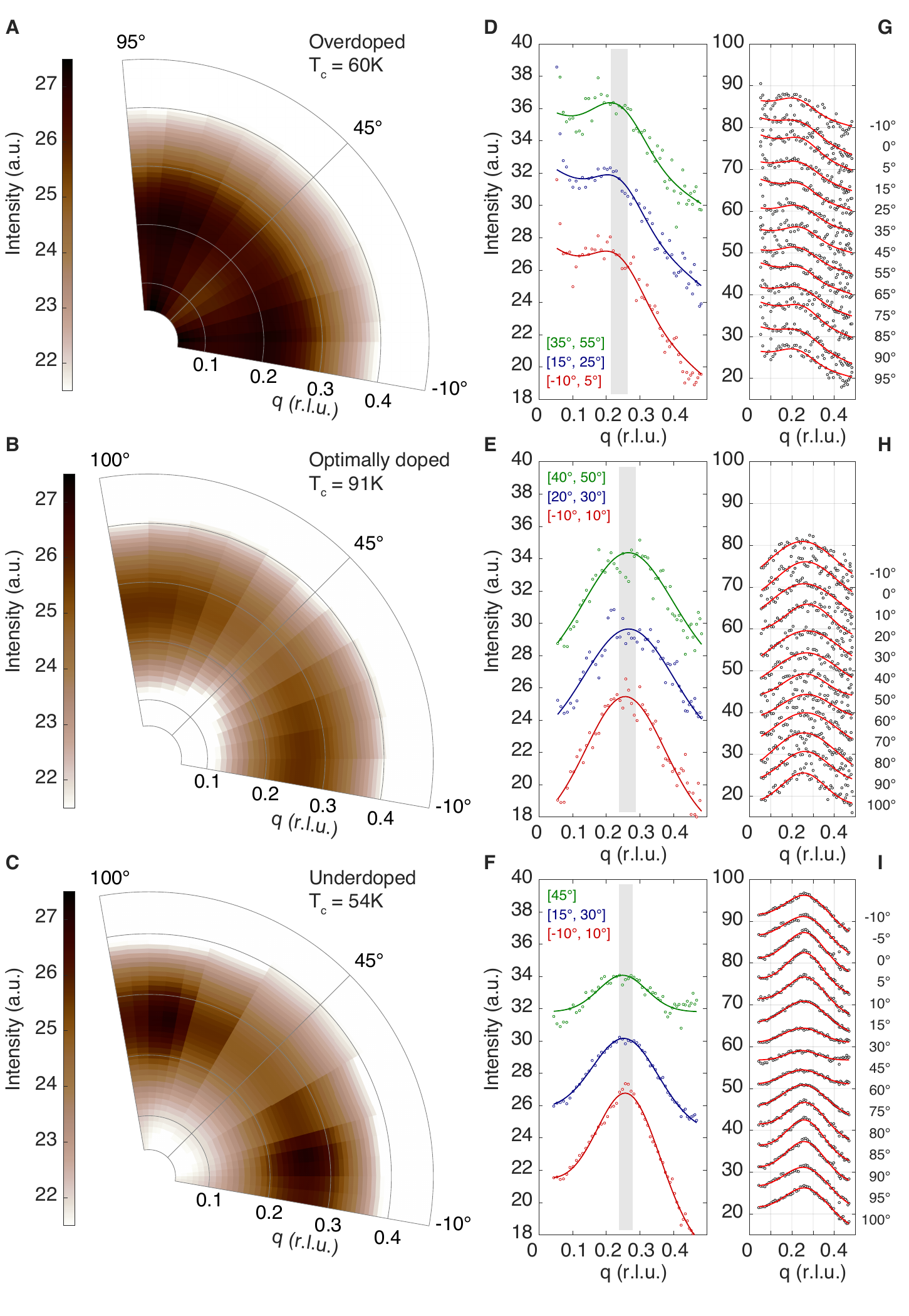}
    \caption{\textbf{Doping dependence from medium resolution RIXS} (\textbf{A-C}) Normalized energy-integrated RIXS mapping showing high energy quasi-circular electron correlations in overdoped, optimally doped and underdoped samples, respectively. (\textbf{D-F}) $q$-cuts integrated over different $\phi$ ranges, as specified in the legends. (\textbf{G-I}) The normalized energy-integrated RIXS data used to used to generate (A and D), (B and E) and (C and F).}
    \label{FSM6}
\end{figure}

\noindent \large{Medium resolution RIXS}\\
\normalsize
In Fig.\,\ref{FSM6} we show the results of medium resolution RIXS done at the qRIXS endstation at the Advanced Light Source in the Lawrence Berkeley National Laboratory. The data were obtained by integrating the RIXS spectra over the $-0.5$ to $0.7$\,eV energy window and normalizing them by spectra integrated over all energies, which allows a comparison between the three different dopings. The data was also symmetrized about the high-symmetry $\phi=45^\circ$ direction. In Fig.\,\ref{FSM6}(\textbf{D}-\textbf{I}) the solid lines are fits of a Gaussian function plus a linear background to the data. The maps in Fig.\,\ref{FSM6}(\textbf{A}-\textbf{C}) were generated from the fits in Fig.\,\ref{FSM6}(\textbf{G}-\textbf{I}), respectively. The gray bars in Fig.\,\ref{FSM6}(\textbf{D}-\textbf{F}) are centered at the average radii of the correlations, obtained from averaging over $\phi$ the peak positions obtained from the fits in Fig.\,\ref{FSM6}(\textbf{G}-\textbf{I}). The widths of the grey bars in Fig.\,\ref{FSM6}(\textbf{D} and \textbf{E}) are obtained from the $95$\% confidence intervals obtained from the fits in Fig.\,\ref{FSM6}(\textbf{G} and \textbf{H}), summing them in quadrature and taking their square root. The same procedure underestimates the uncertainty for the $T_c = 54$\,K. Instead the width of the grey bar in Fig.\,\ref{FSM6}(\textbf{F}) is calculated by taking the lowest and largest peak positions over all $\phi$, taking into account the $95$\% confidence intervals from the fits in Fig.\,\ref{FSM6}(\textbf{I}). The width of the grey bar in Fig.\,\ref{FSM6}(\textbf{F}) The data used to generate Fig.\,\ref{FSM6}(\textbf{C}, \textbf{F} and \textbf{I}) were used in a previous publication [Boschini \textit{et al.} Nat. Comm. \textbf{12}, 1-8 2021]. The new data follows the same experimental procedures as the previously published data, so we direct the reader to [Boschini \textit{et al.} Nat. Comm. \textbf{12}, 1-8 2021] for further details of the experimental procedure.

\newpage

\begin{figure}[H]
    \centering
    \includegraphics[width=0.75\textwidth]{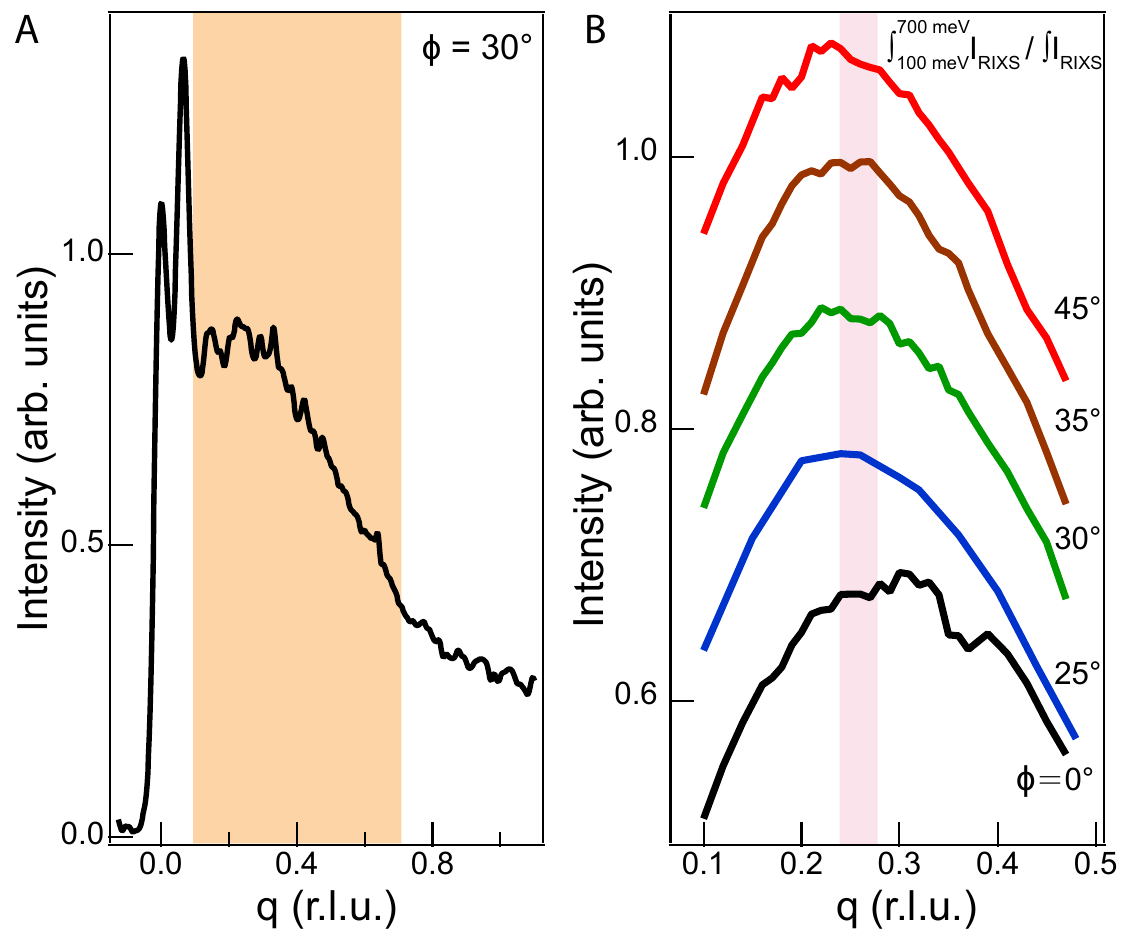}
    \caption{\textbf{Energy-integrated RIXS maps from high resolution RIXS on Bi2212 underdoped T$_{\text{c}}$=54\,K at I21.} \textbf{A} Energy-Loss RIXS spectrum for ($q$,$\phi$)=(0.28\,rlu, 30$^o$). The orange shadow highlights the energy integration window [0.1,0.7]\,eV. \textbf{B} $q$-cuts of the RIXS spectra integrated over the energy regions highlighted in A and normalized by the total energy-integrated RIXS signal. The overall $q$-dependence and position of the maximum is similar to what observed via medium resolution RIXS (see Fig.\,\ref{FSM6}(F and I)). The pink bar reproduces the grey bar in Fig.\,\ref{FSM6}(\textbf{F}), which is obtained from the analysis of the correlations observed with medium resolution RIXS for underdoped Bi2212 (T$_{\text{c}}$=54\,K).}
    \label{FSM7}
\end{figure}

\normalsize

\bibliographystyle{nsfproposal}
\bibliography{bib_lib}


\end{document}